\documentstyle[prl,aps,epsf]{revtex}
\twocolumn
        
\def\avg#1{\langle #1\rangle}

\begin{document}
\draft
\title{Quantum chaotic attractor in a dissipative system}
\author{W. Vincent Liu and William C. Schieve}
\address{Department of Physics
and Center for Statistical Mechanics \& Complex System,\\ The
University of Texas, Austin, Texas 78712}
\maketitle

\begin{abstract}
A dissipative quantum  system is treated here  by coupling it
with a heat bath of  harmonic oscillators. 
Through quantum Langevin equations and Ehrenfest's theorem, 
we establish explicitly the quantum
Duffing equations 
with a double-well potential chosen. 
A quantum noise term appears the only driving force in dynamics.
Numerical studies show that the chaotic attractor exists in this system 
while chaos is certainly forbidden in the classical counterpart. 
\end{abstract}
\pacs{PACS numbers: 05.45.+b,03.65.Sq,05.40.+j,42.50.Lc}

Quantum chaos of Hamiltonian systems has been studied  extensively
\cite{Giannoni,Reichl,Pattanayak,Ashkenazy}.  
By contrast, very little work
has been done in looking at quantum chaos of dissipative
systems. 
Many quantum mechanical systems
 (e.g., SQUID  with Josephson junction, an atom in a cavity of
electromagnetic fields, NMR quantum measurements), however,
 are neither isolated nor Hamiltonian. 
These interact with their environment and
 thus are open in general and noisy and dissipative. 
 Dissipation is relatively difficult to treat in a  quantum system
 since it seems  inevitable to 
deal with stochastic processes via, for instance, the commonly used 
master equation
that is extremely difficult to solve numerically. 
Recently, Spiller and Ralph~\cite{Spiller} studied a damped
 and driven non-linear oscillator by using the quantum state diffusion
 model based on the assumed master equation of Lindblad form \cite{Lindblad}.
They simulated  the behavior of one member of
 the ensemble with a single environment operator and found
that the quantum noise ``kicks'' the motion between the chaotic and
 the periodic behavior and so smears out any fractal structure. 
In general, the quantum state diffusion method
 often demands an approximation of the effects of the complicated environment
by simple operators and always has a problem regarding the
 dimension of the  environment \cite{Schack}.
Brun~\cite{Brun95} has tried to derive the quantum version of 
the forced and damped 
Duffing oscillator through the decoherence approach which involves 
the path integral.  As he noticed,  performing 
calculations in the low-temperature limit is extremely
difficult. For the high temperature limit,
 Brun also suggested  that the quantum
noise ``smears out'' the quantum maps based on Wigner distributions
 and  that numerical computation cannot be
efficiently done because of the necessity of enumerating all the
possible histories and elements of 
the decoherence functional.  

The dissipative quantum  system has  been studied
by means of several different theories, for instance,
 the influence functional
approach of Feynman and Vernon~\cite{Feynman} and its application to
Brownian motion~\cite{Caldeira}, the quantum Langevin
equations~\cite{Gardiner}, the master 
equations~\cite{Spiller,Lindblad,Schack,Gardiner}, etc.~\cite{Parisi}. 
Among  these, the quantum Langevin formalism is more interesting for us
since the dynamics of quantum operators for the system can be
explicitly given in a fashion quite similar to the classical one. 
Most of above theories considered a generic model of the system-bath
interaction.  
Recently,  Pattanayak and Schieve~\cite{Pattanayak}
proposed a derivation of semiclassical dynamics directly from
Heisenberg equations of motion 
via Ehrenfest's theorem where ostensible (apparent) quantum chaos was 
found for the
conservative (Hamiltonian) system with a double-well potential.
In this paper, we will study a dissipative quantum  model similar to
that of Refs.~\cite{Feynman,Caldeira,Gardiner}  and write explicitly 
the quantum Duffing equations for a system of double-well potential
 by means of Langevin formalism
and Ehrenfest's theorem. Numerical results are now possible to obtain
and show 
that chaotic attractors robustly
exist in this model whereas chaos is impossible for
the corresponding classical Duffing system due to the absence of
an external driving force. 

To do this, let us consider a particle of unit mass moving in a one-dimensional
time-independent bounded potential $V(\hat{Q})$ with its Hamiltonian,
$
H_{sys} ={1\over 2}\hat{P}^2 + V(\hat{Q}).
$
To introduce dissipation, we may take the system $H_{sys}$ linearly 
interacting with an
external ``heat bath'' of many degrees of freedom, which here is assumed an
assembly of harmonic oscillators~\cite{Feynman,Caldeira,Gardiner}.  The
familiar example of this model 
is  a system of an atom interacting with a bath of equilibrium photons.
Then the complete Hamiltonian is \cite{Feynman,Caldeira,Gardiner}, 
\begin{eqnarray}
H   &=&	{\hat{P}^2 \over 2} + V(\hat{Q})  +   
	\sum_{n}\left\{  {\hat{p}_n^2 \over 2m} + {1 \over 2}m \omega_n^2
(\hat{x}_n-{C_n\hat{Q} \over m\omega_n^2})^2 \right\}, \label{eq:H}
\end{eqnarray}
in which we have assumed that all harmonic oscillators possess the same
mass $m$ but may have different frequency $\omega_n$ and that $C_n$ is
the coupling constant between the system and the $n$th oscillator. 
The equal times commutation relations implicit
in (\ref{eq:H}) are 
$
{[\hat{Q}, \hat{P}]} = i\hbar, 
{[\hat{Q}, \hat{x}_n]}= {[ \hat{Q}, \hat{p}_n]}=[\hat{P}, \hat{x}_n]=[\hat{P}, \hat{p}_n]=0 , 
{[\hat{x}_l,\hat{x}_n]}=[\hat{p}_l,\hat{p}_n]=0$ and 
${[\hat{x}_l,\hat{p}_n]=i\hbar \delta_{ln}}$.
Now it is straightforward to write down the
Heisenberg equations of motion for the complete system. 
Then we may apply the standard procedure of Ref.~\cite{Gardiner} to derive the
quantum Langevin equations. Assuming a continuous frequency
distribution $g(\omega)$ of 
harmonic oscillators and using the first  Markov approximation
\begin{equation}
{g(\omega){C}^2(\omega) \over m\omega^2} \equiv {2\gamma
\over \pi} \label{eq:markov}
\end{equation}
where $\gamma$ is assumed constant,
we then have
\begin{mathletters}
\label{eq:qdtwo_2}
\begin{eqnarray}
\dot{\hat{Q}} &=&\hat{P}, 	\label{eq:qdtwo_2a}
\\
\dot{\hat{P}}&=& -V^\prime(\hat{Q}) -\gamma \hat{P} + \hat{\xi}(t),
\label{eq:qdtwo_2b} 
\end{eqnarray}
\end{mathletters}
where $\hat{\xi}(t)$ is the quantum noise operator due to the heat bath
and $\gamma$ defined by (\ref{eq:markov})  clearly represents
the constant damping coefficient.
Suppose that the system and the bath are  initially independent (at
$t\rightarrow -\infty$) so that the complete density operator can be
written into 
$\rho=\rho_{sys} \otimes \rho_{b}$, and that the bath is initially
thermal, $\rho_b \sim \exp(-H_b/k_BT)$. Then, 
 $\hat{\xi}(t) $ has the
properties (see, for example, \cite{Gardiner})
\begin{mathletters}
\label{eq:noise}
\begin{eqnarray}
\avg{\hat{\xi}(t)} & =& 0, 	\label{eq:noise_a}\\
  \avg{[\hat{\xi}(t^\prime), \hat{\xi}(t)]_+}& =&  {2\gamma\hbar\over \pi}
 	 \int_0^\infty d\omega \omega \coth({\hbar\omega \over 2k_BT})
 	\cos\omega(t^\prime-t), \label{eq:correlation}
\end{eqnarray}
\end{mathletters}
in which $[\cdots]_+$ denotes an anticommutator. The average
$\avg{\cdots }$ is over all bath variables.
The operator nature of $\hat{\xi}(t)$ can be 
reduced by using the strategy of the adjoint
commutative representation \cite{Gardiner2}. 
We can define a new
operator $\eta(t)$ by $\eta(t) \rho(t^\prime) \equiv {1
\over 2} [\hat{\xi}(t), \rho(t^\prime)]_+ $ for all $t$ and $t^\prime$,
which  yields $[\eta(t),\eta(t^\prime)]=0$. 
This means that $\eta(t)$  is  a $c$-number function of
time.
Let us replace the operator $\hat{\xi}(t)$ by the $c$-number $\eta(t)$
in (\ref{eq:qdtwo_2}). Its
$1$- and $2$-point correlation functions are given by (\ref{eq:noise}).
In the following we will restrict ourselves in
 low temperature limit (not discussed 
in \cite{Brun95}), $T \rightarrow 0$, in which quantum effects are
most important. In this limit the noise $\eta(t)$ is
given by
\begin{mathletters}
\label{eq:noise_lowT}
\begin{eqnarray}
\avg{\eta(t)} & =& 0, 	\\
  \avg{\eta(t) \eta(t^\prime)}& =&{\gamma\hbar\over \pi}
 	 \int_0^\infty d\omega \omega \cos\omega(t^\prime-t) \nonumber
  \\
      &=&
-{\gamma\hbar \over \pi} {1 \over (t-t^\prime)^2}. \label{eq:correlation_num}
\end{eqnarray}
\end{mathletters}
To obtain this, an exponential
cutoff ($e^{-\epsilon\omega}$) was used, letting $\epsilon \rightarrow 0$
after integration \cite{remark_cutoff}.

While the dependence of bath operators has been eliminated in it, 
the quantum Langevin equations (\ref{eq:qdtwo_2}) are still operator
equations and are as difficult as that in the
deterministic quantum mechanics. 
To make further progress,  we make an additional assumption that 
the  wave packet of the system can be described by the
squeezed coherent state~\cite{Perelmov}. Then we have the
relations~\cite{Pattanayak}
$\langle \tilde{Q}^{2m}\rangle =(2m)! (\hbar\mu)^m/m! 2^m, 
\langle \tilde{Q}^{2m+1} \rangle  =0,
 \langle \tilde{P}^2\rangle = \hbar(1+\alpha^2) /4\mu, 
\langle\tilde{Q}\tilde{P} +\tilde{P}\tilde{Q} \rangle =\hbar \alpha,
$
where  $\tilde{O}{=} \hat{O}- \langle \hat{O} \rangle$ henceforth and
$\avg{\cdots}$ denotes the expectation value. It
has been shown that these 
 are exactly equivalent to those derived from the generalized Gaussian wave
functions~\cite{Jackiw,Heller,Tsue}.
The equations of motion for the centroid of a wave packet representing
the  particle are  given from (\ref{eq:qdtwo_2}) by
\begin{mathletters}
\begin{eqnarray}
\avg{\dot{\hat{Q}}}  &=& \avg{\hat{P}}, \\ 
\avg{\dot{\hat{P}}} &=& -\avg{V^\prime(\hat{Q})} -\gamma \avg{\hat{P}}
+ \eta(t).
\end{eqnarray}
\end{mathletters}
We now expand the equations around the centroid by using the identity
$
F(\hat{Q}) = \sum_n  F^{(n)} \tilde{Q}^n /n! $,
where $F^{(n)} =\partial^nF(\hat{Q})/\partial \hat{Q}^n
|_{\hat{Q}=\avg{\hat{Q}}}$, and for a double-well potential of
$V(\hat{Q}) = -{1\over 2} a \hat{Q}^2 + {1\over 4} b \hat{Q}^4$
obtain the closed system of the stochastic differential equations 
\begin{mathletters}
\label{eq:qdo}
\begin{eqnarray}
\dot{Q} & =& P, 	\label{eq:qdo1}\\
\dot{P} &=& aQ -bQ^3 -3bQ\hbar\mu   -\gamma P + \eta(t),
\label{eq:qdo2}  \\
 \dot{\mu} &=& \alpha, \\
\dot{\alpha} &=& {1+\alpha^2 \over 2\mu} + 2\mu(a-3bQ^2) -6b\hbar\mu^2
-\gamma \alpha. \label{eq:qdo4}
\end{eqnarray}
\end{mathletters}
Here we have written $Q,P$ for $\avg{\hat{Q}}, \avg{\hat{P}}$. 
 We remind the reader that all physical observables must be obtained  by
averaging over stochastic noise $\eta(t)$. 
When damping constant $\gamma$ is set to zero, the system
is Hamiltonian  with $H_{extended}={P^2\over 2} +{\pi^2\over 2} -{a\over
2}(Q^2+\rho^2) +{1\over 8\rho^2} + {b\over 4}(Q^4 +3\rho^4+ 6\hbar
Q^2\rho^2)$, for which $\mu\rightarrow \rho^2$ and $\alpha \rightarrow
2\rho\pi$ in (\ref{eq:qdo}). The
above equations of motion can be  reduced to those of
Ref.~\cite{Pattanayak}, in which Hamiltonian semiquantal chaos was
reported. 

Also, in the classical limit $\hbar \rightarrow 0$ 
the first two equations of (\ref{eq:qdo}) decouple from
the fluctuation variables ($\mu,\alpha$) and the noise term 
 vanishes as $T\rightarrow 0$ \cite{Liu}. The 
well-known classical Duffing equations without external 
driving are recovered (for $T\rightarrow 0$) 
\begin{mathletters}
\label{eq:cdo}
\begin{eqnarray}
\dot{Q} & =& P, 	\label{eq:cdo1}\\
\dot{P} &=& aQ -bQ^3   -\gamma P.
\label{eq:cdo2} 
\end{eqnarray}
\end{mathletters}
This shows that (\ref{eq:qdo}) are indeed the 
quantum analog of Duffing equations with the presence of a quantum
noise term, which serves as a Langevin  driving force. 
Now a question perhaps rises upon how effective our
equations (\ref{eq:qdo}) could be to give the description
of exact quantum dynamics.
The procedure of using Gaussian wavepacket to describe the
motion of a particle may be subject to large errors  and will
even break down after a certain time scale
\cite{Pattanayak94b}. This has yet to be decided.  It has been argued,
for example, by Heller \cite{Heller} who has extensively 
studied quantum chaos without noise that the generalized Gaussian
approximation works 
extremely well.  Also, recently Ashkenazy {\it et al}
\cite{Ashkenazy} have  computed the time development
of the wave function in the presence of a potential
barrier in a bounded well for a long time. 
They numerically confirmed the appearance of
Hamiltonian 
chaos due to tunneling first suggested in Ref.~\cite{Pattanayak}. 

\begin{figure}[htbp]
\begin{center}
\leavevmode
\epsfxsize=3in
\epsfbox{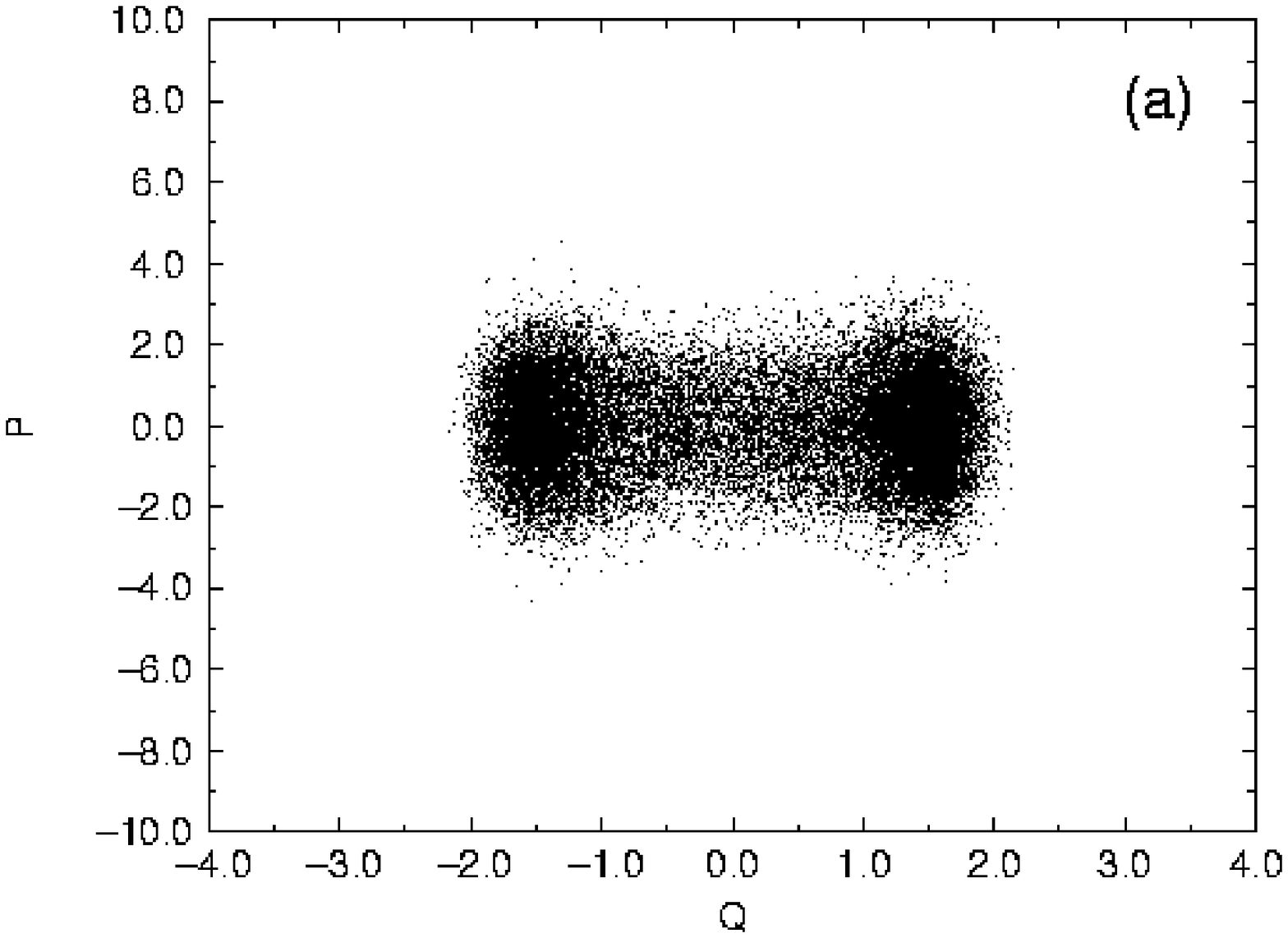}
\epsfxsize=3in
\epsfbox{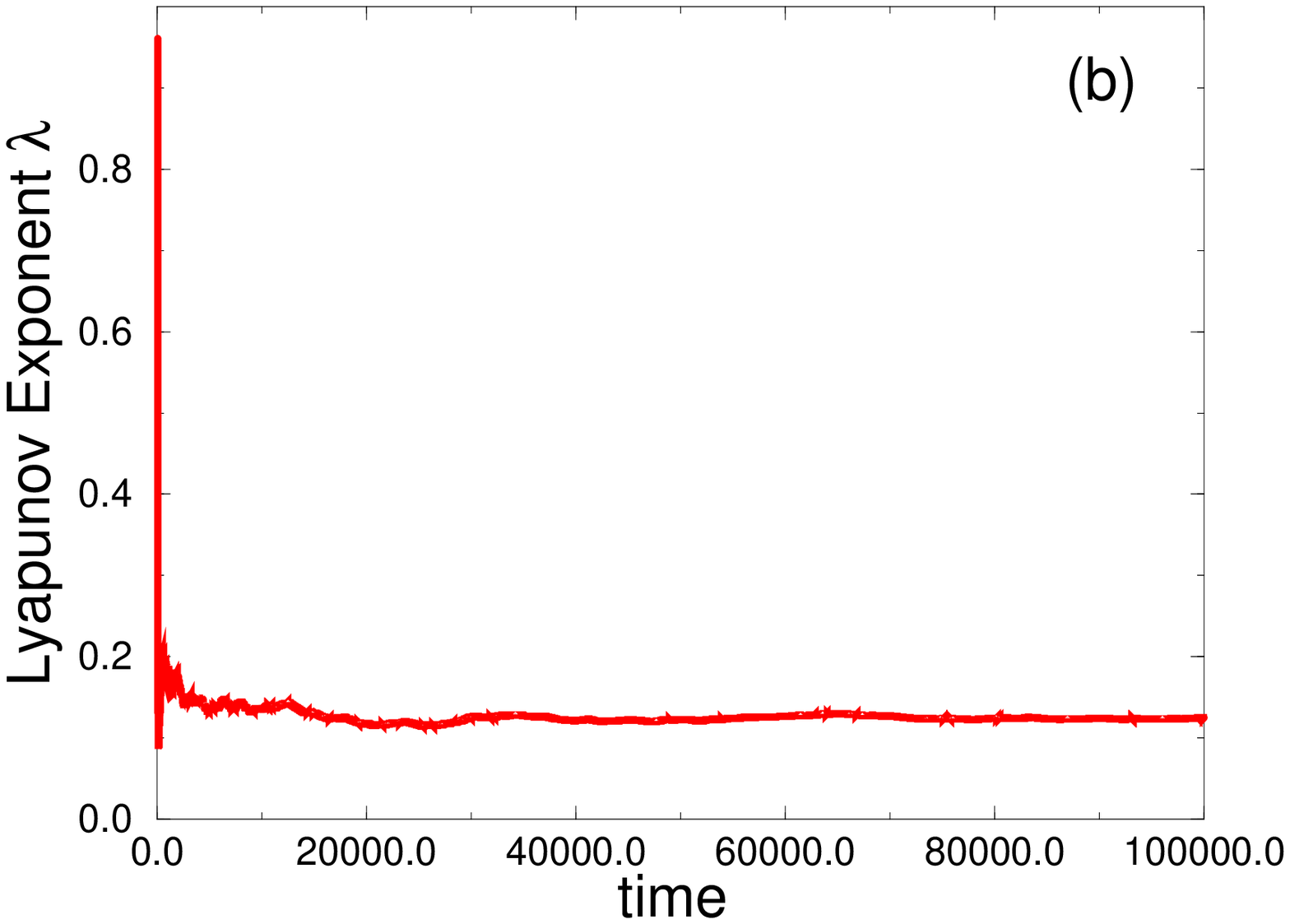}
\end{center}
\caption{Quantum chaotic attractor for one realization with parameters:
$\hbar\equiv 1$, 
$a=10$, $b=4$, $\gamma =0.002$. (a) The Poincar\'e section $Q$-$P$
taken by $\alpha=0\pm 0.0015$ and $\dot{\alpha}\geq 0$; (b) The time
evolution of the largest Lyapunov exponent $\lambda$. For this
realization, the sum of all Lyapunov exponents in  phase space
$\sum_i \lambda_i \simeq -4.011\times 10^{-3}$ and the fractal dimension
$D_f\simeq 3.968$.} 
\label{fig:attractor}
\end{figure}

In order to proceed with simulations of the stochastic differential equations 
(\ref{eq:qdo}), we write \cite{Gardiner} 
$
\eta(t) = \int_{-\infty}^\infty d\omega \, \nu(\omega) \sqrt{|\omega| 
\gamma\hbar\over 2\pi} e^{-i\omega t}$,
where $\nu(\omega) $ is a random function with the properties: 
$\avg{\nu(\omega)}=0,
\avg{\nu(\omega)\nu(\omega^\prime)}=\delta(\omega-\omega^\prime) $,
$\nu(\omega)=\nu(-\omega)$. One
can easily check that $\eta(t) $ generated in this way
satisfies  (\ref{eq:noise_lowT}). It follows that
(\ref{eq:qdo}) can be numerically solved for each realization of the 
random process $\eta(t)$ by the Runge-Kutta method for the  ordinary
differential equations once the random sequence of
$\nu(\omega)$ has been generated.
 Quantum chaotic attractors
are then found for weak damping after some evolution time
(typically around $1000$ in our simulations).
For one realization of
stochastic process $\eta(t)$,  
Fig.~\ref{fig:attractor}(a) shows the structure of the quantum 
chaotic attractor   in the phase
space.  The fact that 
the attractor diffuses out  due to the quantum
noise agrees with
the previous results \cite{Spiller,Brun95}. In particular, one may
compare this attractor with that of Ref.~\cite{Spiller} for the case of
strong noise. 
 Lyapunov exponents $\lambda$s and the fractal dimension
$D_f$ are calculated for each realization by using 
the standard  method as given by Ref.~\cite{Wolf} and, as seen in
Fig.~1(b), are 
saturated after time of the order $t=10000$ with a small residual
oscillation less than $1\%$ ($\lambda_{largest}\simeq 0.124500$).  
For this case, we furthur checked the
largest Lyapunov exponent by using an alternative computation
algorithm \cite{Schuster}  and 
found that the Lyapunov exponents $\lambda_{largest} \simeq 0.124470$
agreeing with the above number up to  an accuracy of $1\%$. 

\begin{figure}[htbp]
\begin{center}
\leavevmode
\epsfxsize=3in
\epsfbox{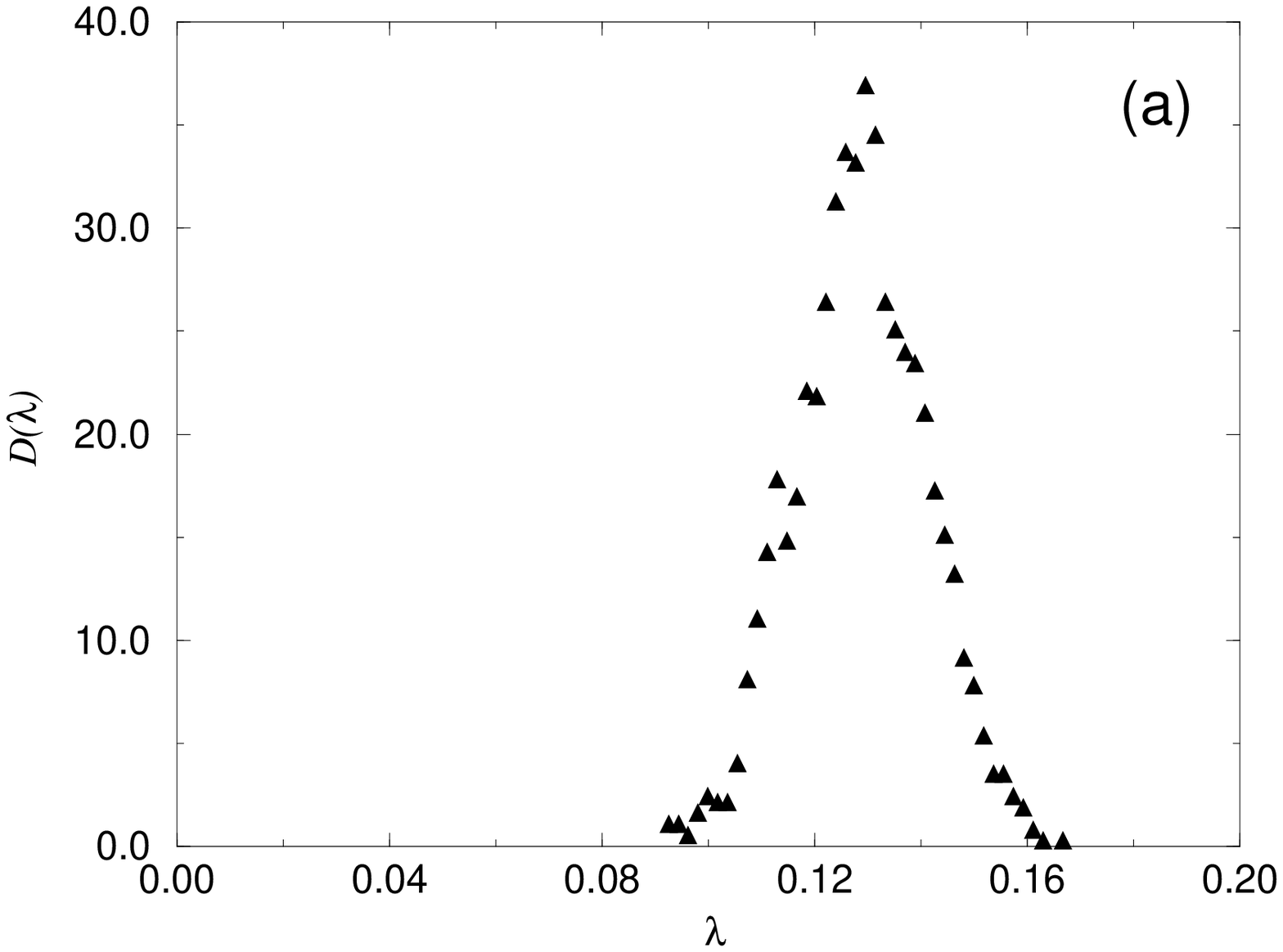}
\epsfxsize=3in
\epsfbox{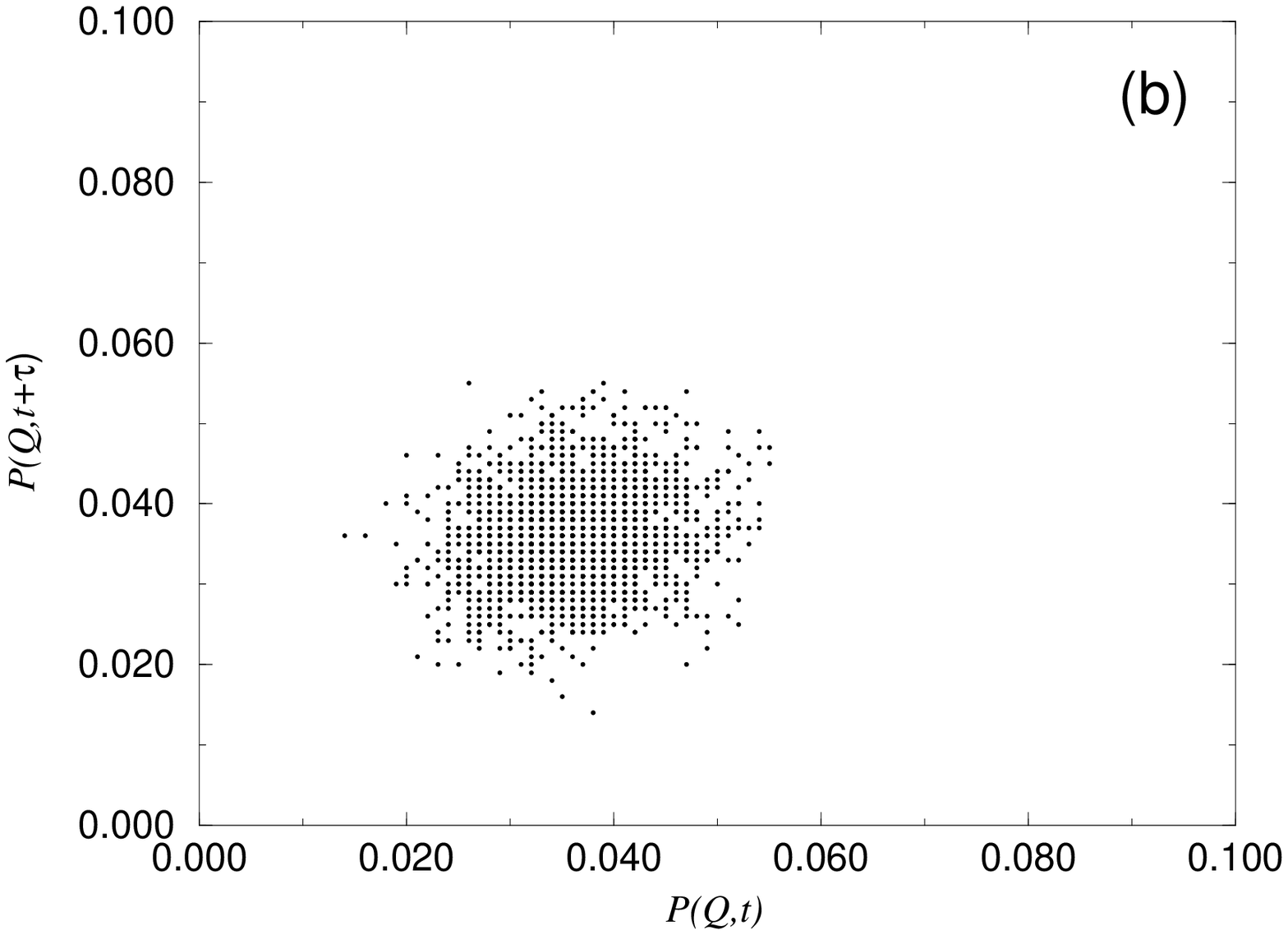}
\end{center}
\caption{The simulation for $1000$ realizations with the same
parameters as in Fig.~\ref{fig:attractor}.
(a) The distribution $D(\lambda)$ of the largest Lyapunov exponents with 
$\int  d\lambda D(\lambda) =1$. 
The mean $\lambda_{\rm largest} \simeq 0.127$, 
   $D_f \simeq 3.969$ and
  $\sum_i\lambda_i \simeq -3.957\times 10^{-3}$.  Lyapunov exponents
are calculated by means of  Wolf {\it et al} \protect\cite{Wolf}.
(b) The probability
map, taking constant $Q=1.581 \pm 
0.05$ (near the bottom of one potential well) and $\tau=5$. This map
starts after a relaxation time of $10000$.} 
\label{fig:distribution}
\end{figure}

We have computed $1000$ samples (realizations)
synchronously and then have obtained the distribution of Lyapunov exponents
$\lambda$ and a probability  map (see Fig.~\ref{fig:distribution}).
As seen in Fig.~\ref{fig:distribution}(a), the largest 
Lyapunov exponents for all
realizations  are conclusively positive.
Note that the fractal dimension $D_f$ is very close to four due to the weak
damping ($\gamma=0.002$) since $D_f \rightarrow 4$ as $\gamma
\rightarrow 0$ where the system becomes Hamiltonian. 
Now let ${\cal P}(Q,t)$ denote the probability of the system at
position $Q$ at time $t$. We make use of a
probability map \cite{Kapitaniak} defined by ${\cal P}(Q,t) \rightarrow {\cal
P}(Q,t+\tau)$  for constant $Q$ and $\tau$  
to further characterize the behavior of the noisy quantum system. 
It has been shown \cite{Kapitaniak} that 
for regular behavior the
probability  function ${\cal P}(Q,t)$ does not depend on time
once the motion of the system is stable 
and therefore the map ${\cal P}(Q,t) \rightarrow {\cal P}(Q,t+\tau)$ should
only consist of a single point. 
Hence,
Fig.~\ref{fig:distribution}(b)  
justifies the chaotic behavior of this noisy quantum  system as well
\cite{remark_prob}.  
By contrast, it is well  known that for the corresponding classical Duffing
equations (\ref{eq:cdo})
only point attractors,  describing regular
motions, can be allowed 
 due to the absence of the external driving force.   
While it has been suggested \cite{Pattanayak,Ashkenazy} that the
tunneling effect induces chaotic
behavior in a  Hamiltonian quantum system,   the dynamical effect of
the quantum noise in (\ref{eq:qdo})  should have played 
 a crucial role to form the stable chaotic attractor in our dissipative
quantum system. Otherwise, the system has to
be attracted to the bottom of either well with zero momentum.  
In other words, we report here that it is the quantum noise that leads to the 
 chaotic attractor. 
Indeed, this is somewhat reminiscent of the classical fact that either
multiplicative or additive noise may induce homoclinic crossing and so
chaos,  as suggested by Schieve, Bulsara and Jacobs \cite{Schieve} 
 in their studies for  classical
stochastic chaos.

To conclude we note  that the quantum Duffing equations 
in low temperature limit have been
explicitly established from the quantum Langevin equations.   
These equations manifestly display great advantage of numerical
computation in comparison with others \cite{Spiller,Brun95}. 
Numerical results 
show that these equations exhibit the stable chaotic
attractor  for weak damping 
while as already known
the classical counterparts certainly forbid chaos due to the absence of
an external driving force. To our knowledge, this is the
first study of the  dynamical behavior of the dissipative
quantum system without external driving showing the quantum chaotic
attractor for such a system. More detailed studies shall be presented
elsewhere~\cite{Liu}. 

We should bear in mind that some assumptions  have been made.
First, all results are
based on a simplified theoretical system-plus-environment model for 
which  we assumed that: $i)$
the heat bath consists of an assembly of harmonic oscillators; $ii)$
there is a continuous distribution of oscillator frequencies;
$iii)$ the coupling of the system  to the bath operators is
linear and the coupling constant is a smooth function of oscillator
frequency; and $iv) $ the stochastic process is Markovian.  All these are
quite well-known and usual in the study of 
a dissipative quantum system (see, for example,
Refs.~\cite{Feynman,Caldeira,Gardiner}). 
Second, the squeezed coherent state (or equivalently the generalized
Gaussian wave-packet) has been used to approximate the true wave function of
the system. The full quantum phase space is thus restricted
into a truncated ``semiquantal'' phase space \cite{Pattanayak,Pattanayak94b}.
It has been shown by Ashkenazy {\it et al.} \cite{Ashkenazy}
  via computer simulation that this approximation does
not break down for a  long time.
Fully understanding its validity  is still an open task.

The authors would like to thank Arjendu K. Pattanayak for many
stimulating discussion during his recent visit.

\twocolumn


\begin{thebibliography}{99}
\bibitem{Gardiner2} This technique with its applications was
discussed in Sects. 3.5, 3.7  and 9.3 of Ref.~\cite{Gardiner}.
\bibitem{remark_cutoff} This was given in
Ref.~\cite{Gardiner}. Since the high frequency behavior of the
spectral function  is proportional to
$|\omega|$ and so is very badly behaved in noise theory, we have to
impose a high frequency cutoff by introducing
$e^{-\epsilon\omega} (\epsilon \ll 1)$.
The physical justification for this
comes from the form (\ref{eq:markov}),  
where the factor $g(\omega) {C}^2(\omega)$ comes into play.
This factor must drop off sufficiently rapidly at high frequencies
because for wavelengths shorter than $\lambda_{\rm cutoff}=2\pi c\epsilon$  a
wave of displacements of atoms  becomes meaningless.
\bibitem{remark_prob} 
Because we have only run $1000$ realizations (so the accuracy
of ${\cal P}(Q,t)$ cannot be better than $0.001$), 
the map seems to contain only finite elements.  We argue that for  an
infinite number of realizations (i.e., in the 
thermodynamical limit) the map shall consist of infinite number of
points.
\end{thebibliography}
\end{document}